\documentstyle[12pt]{article}
\begin{document}
\centerline{\bf Dynamics of chaotic inflation with variable space dimension}
\vspace*{0.050truein}
\centerline{Forough Nasseri\footnote{E-mail:
nasseri@shahrood.ac.ir}}
\centerline{\it Department of Physics, Shahrood University,}
\centerline{\it P.O.Box 36155-316, Shahrood, Iran}
\centerline{\it Institute for Studies in Theoretical Physics
and Mathematics,}
\centerline{\it P.O.Box 19395-5531, Tehran, Iran}
\begin{center}
(\today)
\end{center}

\begin{abstract}

Within the framework of a model Universe with variable space dimension,
we study chaotic inflation with the potential $m^2\phi^2/2$, and calculate
the dynamical solutions of the inflaton field, variable space
dimension, scale factor, and their interdependence during the
inflationary epoch. We show that the characteristic of the variability
of the space dimension causes the inflationary epoch in the variable space
dimension to last longer than the inflationary epoch in the constant
space dimension.

\noindent
PACS number(s): 98.80.Cq, 04.50.+h, 98.80.Bp, 98.70.Vc\\
Preprint: IPM/P-2001/048
\end{abstract}  

\section{INTRODUCTION}

Interest in a speculative model in which the number of space dimension
decreases continuously as the Universe expands has increased during the 
past few years [1-5].

A model Universe with variable space dimension proposed in \cite{4}
seems to be singularity free, having two turning points for the space 
dimension. The way of generalizing the standard cosmological model to
the variable space dimension used in \cite{4} is questioned,
and another way of writing the field equations is proposed \cite{2}. 
It has been pointed out that the model discussed in \cite{4}
has no upper bound for the space dimension.

Authors of Ref.\cite{2} studied critically previous works in 
\cite{4}, and derived new Lagrangians and field equations. 
They also discussed the model Universe with variable space dimension
from the view point of quantum cosmology, and obtained a general wave
function. In the limit of constant space dimension, the wave function of
the model approaches the tunneling Vilenkin wave function or
the modified Linde wave function \cite{2}.

Ref. \cite{1} considers chaotic inflation in the context of a model
with time-varying spatial dimension. The dimensionality of the space is
defined by constructing the space out of ``cells'' with characteristic
dimension $\delta$ crumpled in such a way as to exhibit an
effective dimension $D$.
The authors generalized chaotic inflation with the potential
$m^2 \phi^2/2$ from the three-space to the constant $D$-space, and then to
the variable $D$-space. They also generalized the slow-roll
approximation. After doing some necessary treatments, step by step,
they obtained dynamical equations of the inflaton field,
the scale factor and the variable space dimension.

In Ref. \cite{1}, an upper limit for the space dimension at the
Planck length has been obtained. Based on the radius of the Universe
at the onset of inflation to be larger than the Planck length $l_P$,
and than the minimum size of the Universe $\delta$, the space dimension
at the Planck length must be smaller than or equal to $4$.
This result corresponds to the value of the universal constant of
the model $C$ to be larger than or equal to $\sim 1678.8$.

In this article, we emphasize on a currently active area of research
from the point of view of quantum field theory. In such a research,
length scales of order the current horizon size could very easily have
started out during inflation smaller than the Planck length.
The trans-Planckian problem of inflationary cosmology with variable space
dimension is interesting and worthy of study. We will address this
problem in \cite{f}.

Using the effective time variation of $G$, one can 
conclude an upper limit, to be equal to $4$, for the space dimension at
the Planck epoch. This result corresponds to a lower limit on the value
of the universal constant, $C$, of the model, see Ref. \cite{3}.

A short review of the cosmological model with variable space dimension
is given in Sec.2. Using the previous work in \cite{1}, we obtain in Sec.3.1 
the dynamical solutions of the inflaton field,
the variable space dimension, and the scale factor.
We also calculate in Sec.3.2 how long the inflationary epoch lasts,
and show that the characteristic of
the variability of the space dimension causes the inflationary
epoch in the variable space dimension to last longer than the inflatinary
epoch in the constant space dimension. Sec.4 contains our conclusions.
We will use a natural unit system in which
$c=\hbar=1$ and $l_P=M_P^{-1}=\sqrt{G}=1.6160 \times 10^{-33} {\rm cm}$.

\section{REVIEW OF MODEL UNIVERSE WITH VARIABLE SPACE DIMENSION}

In \cite{{1},{2}}, there are some clarifications about the idea of variable 
space dimension, and some motivations for choosing a model Universe with 
variable space dimension. There are also some reasonably raised questions:
what causes the number of spatial dimension to decrease as the Universe 
expands? Is there any physical process causes or necessitates such a decrease?
And how does this ``disappearance'' of spatial dimensions take place? 
Is it that gravity gradually stops ``feeling'' some dimensions or that the 
size of some of the dimensions shrinks to zero? The reader can find the 
answers of these questions in detail in Refs. \cite{{1},{2}}. It is worth  
mentioning that for generalization of Friedmann equations from three-space
to variable space dimension, we cannot define a metric approach. 
Rather, we use a Lagrangian approach. There is a constraint in this model 
which can be written as  
\begin{equation}
\label{1}
\left( \frac{a}{\delta} \right)^D= \left( \frac{a_0}{\delta} \right)^{D_0}=
e^C,
\end{equation}
or 
\begin{equation}
\label{2}
\frac{1}{D}= \frac{1}{C} \ln \left( \frac{a}{a_0} \right) + \frac{1}{D_0}.
\end{equation}
Here $a$ is the scale factor of Friedmann Universe, $D$ the variable 
space dimension, $\delta$ the characteristic minimum length of the model,
$C$ the universal constant of the model. The zero subscript in any
quantity, e.g. in $a_0$ and $D_0$, denotes its present value. 
In our formulation, we consider the comoving length of the Hubble radius
to be equal to one. So the scale factor means the physical length
in cosmology which is the Hubble radius.
Note that in Eqs. (\ref{1}) and (\ref{2}), the space dimension is  
a function of cosmic time $t$. Time derivative of Eq. (\ref{1}) leads to 
\begin{equation}
\label{3}
\dot{D}=-\frac{D^2 \dot{a}}{C a}.
\end{equation}
It is easily seen that the case of constant space dimension corresponds to
the limit of $C \to +\infty$. Substituting the Ricci scalar for a
$D$-manifold with constant space dimension, and constant curvature
$k=-1, 0, +1$
\begin{equation}
\label{4}
R= \frac{D}{N^2} \left\{ \frac{2 \ddot{a}}{a} + (D-1) \left[ 
\left( \frac{\dot a}{a} \right)^2 + \frac{N^2 k}{a^2} \right]
- \frac{2 \dot{a} \dot{N}}{a N} \right\},
\end{equation}
in the Einstein- Hilbert action for pure gravity
\begin{equation}
\label{5}
S_G=\frac{1}{2 \kappa} \int d^{(1+D)} x \sqrt{-g} R,
\end{equation}
and using Hawking-Ellis action of a perfect fluid, for the model Universe 
with variable space dimension, the following Lagrangians have been
obtained  
\cite{2}
\begin{eqnarray}
\label{6}
{\cal{L}}_I &:=& -\frac{V_D}{2 \kappa N} \left( \frac{a}{a_0} \right)^D
D(D-1) \left[ \left( \frac{\dot{a}}{a} \right )^2 - \frac{N^2 k}{a^2}
\right]\nonumber\\
&-& \rho N V_D \left( \frac{a}{a_0} \right)^D,
\end{eqnarray}
and
\begin{eqnarray}
\label{7}
{\cal{L}}_{II} &:=& - \frac{V_D}{2 \kappa N} 
\left( \frac{a}{a_0} \right)^D\nonumber\\
& \times & \left( \frac{2 \dot{D} \dot{a}}{a} + \frac{2 D \dot{a} \dot{D}}{a}
\ln \frac{a}{a_0} + D(D-1) \left[ \left( \frac{\dot{a}}{a} \right)^2
-\frac{N^2 k}{a^2} \right] \right) \nonumber\\
&-& \frac{1}{\kappa N} \frac{D \dot{D} \dot{a}}{a} \frac{d V_D}{dD}
\left( \frac{a}{a_0} \right)^D - \rho V_D N \left( \frac{a}{a_0}
\right)^D,
\end{eqnarray}
where $\kappa=8 \pi M_P^{-2}$, $N$ the lapse function, $\rho$ the energy 
density, and $V_D$ the volume of the space-like sections 
\begin{eqnarray}
\label{8}
V_D &=& \cases {\frac{2 \pi^{(D+1)/2}}{\Gamma[(D+1)/2]},          
               & if $\;k=+1$, \cr          
               \frac{\pi^{(D/2)}}{\Gamma(D/2+1)}{\chi_c}^D,          
               & if $\;k=0$, \cr
               \frac{2\pi^{(D/2)}}{\Gamma(D/2)}f(\chi_c),            
               & if $\;k=-1$. \cr}
\end{eqnarray}
Here $\chi_c$ is a cut-off, and $f(\chi_c)$ is a function thereof,
see Ref. \cite{2}. In Eq.(\ref{4}), Ricci scalar has been given 
for an arbitrary constant space dimension. To get the action of the model, 
one can substitute Eq.(\ref{4}) in Eq.(\ref{5}), and consider the space 
dimension as a function of cosmic time. Depending on the way of
integration in Einstein-Hilbert action, 
Lagrangian ${\cal{L}}_I$ and ${\cal{L}}_{II}$
can be obtained. In the limit of constant space dimensions, or $D=D_0$ and 
$C \to +\infty$, ${\cal{L}}_I$ and ${\cal{L}}_{II}$ approach to 
Einstein-Hilbert Lagrangian which is
\begin{eqnarray}
\label{9}
{\cal{L}}^0_{I,II} &:=& - \frac{V_{D_0}}{2 \kappa_0 N}
\left( \frac{a}{a_0} \right)^{D_0} D_0 (D_0-1) \left[ \left( \frac{\dot a}{a}
\right) ^2 - \frac{N^2 k}{a^2} \right]\nonumber\\
&-& \rho N V_{D_0} \left( \frac{a}{a_0} \right)^{D_0}.
\end{eqnarray}
So, Lagrangian ${\cal{L}}_I$ and ${\cal{L}}_{II}$ cannot abandon Einstein's 
gravity. We have introduced $\kappa_0$ for the value of the gravitational 
coupling constant in the case of the constant space dimension, 
$D=D_0$ \cite{2}. It is worth mentioning that the model presented here
is consistent in the limit of a flat spacetime. This is very important
when considering the overall viability of the model.
In the limit $k=0$, the Lagrangian ${\cal{L}}^0_{I,II}$ leads to
the flat FRW cosmology. Varying the Lagrangian ${\cal{L}}_I$ with respect
to $N$ and $a$, we find the following equations of motion in the gauge
$N=1$, respectively
\begin{eqnarray}
\label{10}
&&\left( \frac{\dot a}{a} \right)^2 + \frac{k}{a^2} = 
\frac{ 2 \kappa \rho}{D(D-1)},\\
\label{11}
&&(D-1)\left\{ \frac{\ddot a}{a} + \left[ \left( \frac{\dot a}{a}
\right)^2 +\frac{k}{a^2} \right] \left( - \frac{D^2}{2C} \frac{d \ln
V_D}{dD} -1 - \frac{D(2D -1)}{2C(D-1)} + \frac{D^2}{2D_0}
\right) \right\}\nonumber\\
&&+ \kappa p \left( - \frac{d \ln V_D}{dD} \frac{D}{C} - \frac{D}{C} \ln
\frac{a}{a_0} +1 \right)=0.
\end{eqnarray}
The continuity equation of the model Universe with variable space dimension
can be obtained by (\ref{10}) and (\ref{11}) 
\begin{equation}
\label{12}
\frac{d}{dt} \left[ \rho \left( \frac{a}{a_0} \right)^D V_D \right]+
p \frac{d}{dt} \left[ \left( \frac{a}{a_0} \right)^D V_D \right]=0.
\end{equation}
In the limit of constant space dimension, Eqs.(\ref{10}), (\ref{11})
and (\ref{12}) approach the corresponding equations in the constant 
$D_0$-space \cite{{2},{3}}. A complete discussion of the field equations of 
${\cal{L}}_{II}$ has been given in \cite{2}.
Using (\ref{1}), (\ref{3}) and (\ref{10}), one can get the evolution
equation of the space dimension:
\begin{equation}
\label{13}
{\dot D}^2=\frac{D^4}{C^2} \left[ \frac{2 \kappa \rho}{D(D-1)}
-k \delta^{-2} e^{-2C/D} \right].
\end{equation}
To study chaotic inflation with the potential $m^2 \phi^2/2$
in the framework of the model Universe with variable space dimension, the 
crucial equations are obtained by substituting 
\begin{eqnarray}
\label{14}
\rho \equiv \frac{1}{2}\left( {\dot \phi}^2 + m^2 \phi^2 \right),\\
\label{15}
p \equiv \frac{1}{2} \left( {\dot \phi}^2 - m^2 \phi^2 \right),
\end{eqnarray}
in Eqs.(\ref{10}), (\ref{12}) and (\ref{13}). So, using the slow-roll 
approximations which are 
\begin{equation}
\label{16}
{\dot \phi}^2 \ll m^2\phi^2,\;\;{\ddot{\phi}}\ll D H {\dot{\phi}},\;\;
-{\dot H}\ll H^2,
\end{equation}
we are led to
\begin{eqnarray}
\label{17}
&&\left( \frac{\dot a}{a} \right)^2 \simeq 
\frac{8 \pi m^2 \phi^2}{D(D-1)M_P^2},\\
\label{18}
&&\frac{D^2 \dot{a} \dot{\phi}}{a} \left[ \frac{1}{D_0} -
\frac{1}{C} \left\{\ln\chi_c +\frac{1}{2} \ln \pi
-\frac{1}{2} \psi \left( \frac{D}{2}+1 \right)\right\}\right]
\simeq -m^2 \phi,\\
\label{19}
&&{\dot D}^2 \simeq \frac{8 \pi D^3 m^2 \phi^2}{C^2 (D-1) M_P^2}.
\end{eqnarray}
Note that in (\ref{16}), $D$ is a function of time, and 
$H \equiv {\dot{a}}/{a}$ \cite{1}.
One can rewrite Eq. (\ref{18}) by neglecting the term of the order
${\cal{O}}(C^{-1})$. So, we have
\begin{equation}
\label{18b}
\frac{D^2 \dot{a} \dot{\phi}}{a D_0} \simeq -m^2 \phi.
\end{equation}
Using Eqs. (\ref{3}), (\ref{17}) and
(\ref{18b}), it is easily shown that
\begin{eqnarray}
\label{18h}
\left( \frac{\phi}{M_P} \right)^2 &=&
\frac{C D_0}{4 \pi} \int^D_{D_i}
\frac{dx (x-1)}{x^3} +\left( \frac{\phi_i}{M_P} \right)^2 \nonumber\\
\label{18c}
&=& \frac{CD_0}{4 \pi} \left[ \frac{1}{D} \left( \frac{1}{2D} -1 \right)
-\frac{1}{D_i} \left( \frac{1}{2D_i}-1 \right) \right] +
\left( \frac{\phi_i}{M_P} \right)^2,
\end{eqnarray}
where subscript $i$, for example in $D_i$ and $\phi_i$, denotes its  
value at the onset of inflation, when $t=t_i$ \cite{1}.

As shown in \cite{1}, for different values of $C$, corresponding to
the space dimension at the Planck length  
$D_P = 4, 10, 25$ and $+\infty$, we calculated the e-folding number
${\cal{N}} \sim 69, 98, 116$ and $132$, respectively.
Using the number of e-foldings, we have also obtained the size of the
Universe at the onset of inflation, $a_i$, see Table 1.

\begin{table}
\begin{center}
\caption{Values of $D_P$, $C$, $\delta$, $\cal{N}$, $D_i$, and $a_i$
\cite{1}.}
\begin{tabular}{cccccc} \hline\hline
$D_P$ & $C$ & $\delta({\rm cm})$ & $\cal{N}$ & $D_i$ & $a_i({\rm cm})$
\\ \hline
$3$ & $+\infty$ & $0$& $58.32$ & $3.00$ & $1.02 \times 10^{-25}$ \\
$4$ & $1678.8$ & $8.6 \times 10^{-216}$ & $69.80$ & $3.94$ & $1.06 \times
10^{-30}$ \\
$10$ & $599.57$ & $1.5 \times 10^{-59}$ & $98.97$ & $16.08$ &
$2.28 \times 10^{-43}$ \\
$25$ & $476.93$ & $8.4 \times 10^{-42}$ & $116.42$ & $-22.65$ &
$6.02 \times 10^{-51}$ \\
$+\infty$ & $419.70$ & $l_P$ & $132.25$ & $-7.50$ & $8.03 \times 10^{-58}$
\\ \hline\hline
\end{tabular}
\end{center}
\end{table}

\begin{table}
\begin{center}
\caption{Values of $D_f$, $\phi_i$, and $\phi_f$ \cite{1}.}
\begin{tabular}{ccccc} \hline\hline
$D_P$ & $C$ & $D_f$ & $\phi_i/M_P$ & $\phi_f/M_P$
\\ \hline
$3$ & $+\infty$ & $3.00$& $3.060$ & $1/{\sqrt{4\pi}}$ \\
$4$ & $1678.8$ & $3.38$ & $3.491$ & $0.290$ \\
$10$ & $599.57$ & $4.40$ & $4.512$ & $0.304$ \\ \hline\hline
\end{tabular}
\end{center}
\end{table}

In a standard picture of chaotic inflation, the inflationary period
naturally lasts for many more than $60$ or so e-folds, and the Universe
is deriven exponentially toward flatness, so that $\Omega_0$ is expected
to be unity to high accuracy \cite{7}.

As discussed in \cite{1}, constraints on the dimensionality of the
space are derived from requiring that the radius of the space at the onset
of inflation, $a_i$, be larger than the Planck length, and than the minimum
size of the model, which is $\delta$.
Particularly, for $D_P=25$ and $+\infty$, we have $a_i < \delta < l_P$,
and for $D_P = 10$, $\delta < a_i < l_P$. 
For $D_P=4$, we have $\delta < l_P < a_i$, see Table 1.
The value of the space dimension at the end of inflationary epoch, $D_f$,
and the value of the inflaton field at the onset and end of inflation,
$\phi_i$ and $\phi_f$ are given in Table 2. It should be noticed
that the values of Table 2 are given for $D_P=3, 4, 10$ and not
for $D_P=25$ or so. Since for $D_P \geq 25$ we have $a_i < \delta < l_P$,
we rule out $D_P \geq 25$ \cite{1}.

We should here emphasize on a currently active area of
research from the point of view of quantum field theory.
In such a research, length scales of order the current horizon
size could very easily have started out during inflation smaller
than the Planck length, with no inconsistency from the standpoint of
classical physics [10-22]. Based on the issue of the trans-Planckian
problem some arguments in the paper \cite{1} no longer apply,
particularly the constraints from requiring the initial radius
of curvature to be larger than the Planck length $l_P$.
More clearly, for $D_P=10$ the value of $a_i < l_P$, and
in \cite{1}, we ruled out $D_P=10$. This result of Ref. \cite{1}
could not be correct because for $D_P=10$ we should study in the
trans-Planckian physics. This point will be further discussed in \cite{f}.

We should here emphasize on the meaning of the scale factor
in the model. This is necessary because the analysis depends on an
interpretation scale factor of the Universe as an effective
radius of the space, so that in $D$ dimension, the volume of the
Universe is given by $a^D$, resulting in a constraint on the radius
compared to a fundamental length $\delta$ of the form Eq.(\ref{1}).
Constraints on the dimensionality of the space are
derived from requiring that the radius of the space at the onset of
inflation, $a_i$, be larger than the Planck length. This is a reasonable
procedure in a non-flat ($k= \pm1$), since the scale factor defines a
radius of curvature for such spacetimes, and thus has a straightforward
interpretation as a physical length. This however becomes problematic in
flat ($k=0$) cosmology, since the radius of curvature of the space becomes
infinite, and the scale factor is arbitrary.
The proper or physical length are obtained from the comoving length
by multiplication of the scale factor \cite{7}
\begin{equation}
\label{phys1}
\ell_{physical} = a(t) \ell_{comoving}.
\end{equation}
While the comoving length does not change with time, the proper length
changes with time because of $a(t)$.
We take the scale factor having the dimension
of length, and the comoving length is a dimensionless quantity.
The comoving length is measured by a set of constant rulers, while the
proper length is measured by a set of expanding or contracting rulers.
The flat model is unbounded with infinite volume, and with infinite radius.
For the flat model the scale factor does not represent any physical radius
as in the closed case, or an imaginary radius as in the open case, but
merely represents how the physical distance between comoving points
scales as the space expands or contracts.
Based on Friedmann Universe, for the falt Universe with radiation dominated
\begin{equation}
a(t) = a_0 \left( \frac{t}{t_0} \right)^{1/2},
\end{equation}
and with dust dominated
\begin{equation}
a(t) = a_0 \left( \frac{t}{t_0} \right)^{2/3}.
\end{equation}
So the physical length to the particle or causal
horizon $d_H(t)$, at time $t$ is simply obtained by \cite{subir} 
\begin{eqnarray}
\label{phys3}
d_H(t)=a(t) \int^{t}_0\frac{dt'}{a(t')} = \cases {2t,          
               & radiation dominated, \cr
               3t,
               & dust dominated. \cr}
\end{eqnarray}
In this paper, we take the comoving length of the particle horizon
to be equal to one
\begin{equation}
\ell_{comoving}=\int^{t}_0\frac{dt'}{a(t')}=1.
\end{equation}
This means that $d_H(t)=a(t)$. In this case, the scale factor of the
flat Universe is the physical length, the horizon.

\section{DYNAMICS OF THE MODEL}

Let us now study the dynamical behavior of the inflationary cosmology
with variable space dimension. In  Sec. 3.1, we obtain approximately
the dynamical solution of $\phi(t)$, $a(t)$, and $D(t)$.
In Sec. 3.2, we then study how long the inflatinary epoch lasts.
We show that the inflationary epoch of the model with variable
space dimension is longer than that of the constant space dimension.

\subsection{Dynamical solutions}

Let us now obtain the evolution equations of $\phi(t)$,
$a(t)$ and $D(t)$. Solving Eq. (\ref{18c}), one can get $D$ in terms 
of $\phi$. This relation is given by 
\begin{equation}
\label{21}
D(\phi)= \frac{1+ \sqrt {1-\frac{2}{D_i} \left( 1 - \frac{1}{2D_i} \right)
- \frac{8 \pi}{CD_0} \left[ \left( {\phi_i}/{M_P} \right)^2
-\left( {\phi}/{M_P} \right)^2 \right]}}
{\frac{2}{D_i} \left( 1 -\frac{1}{2D_i} \right) 
+\frac{8 \pi}{CD_0} \left[ \left( {\phi_i}/{M_P} \right)^2
- \left( {\phi}/{M_P} \right)^2 \right] }.
\end{equation} 
It is worth mentioning that Eq.(\ref{21})
approaches $D=D_i$ in the limit of constant space dimension or
$C \to +\infty$.
Using Eqs.(\ref{17}) and (\ref{18b}), the classical
equation of motion for $\phi(t)$ is given by
\begin{equation}
\label{22}
\frac {\dot{\phi}(t)}{M_P} = - \frac{m D_0}{2D}\sqrt{\frac{D-1}{2 \pi D}}.
\end{equation}
This just tells us that the inflaton field is like a ball rolling 
down a hill, $\dot{\phi} <0$. Substituting Eq. (\ref{21}) in 
(\ref{22}), we are led to the non-linear differential equation  
$\dot \phi \equiv f(\phi)$. We approximate this differential equation by
expanding $f(\phi)$ in inverse powers of $C$. We are therefore led to
\begin{equation}
\label{23}
\frac{\dot\phi(t)}{M_P}=\alpha +\frac{\beta}{C}
\left[ \left( \frac{\phi}{M_P}  \right)^2 - \left( \frac{\phi_i}{M_P}
\right)^2 \right] + {\cal{O}}\left( \frac{1}{C^2} \right) + ...,
\end{equation}
where
\begin{eqnarray}
\label{23a}
\alpha &\equiv& - \frac{mD_0}{2 D_i} \sqrt{\frac{D_i-1}{2 \pi D_i}},\\
\label{23b}
\beta &\equiv& \frac{m (2 D_i -3)}{(D_i -1)} \sqrt{\frac{\pi D_i}
{2 (D_i -1)}}.
\end{eqnarray}
The above differential equation is the first order Riccati equation
\cite{solve1}. It can be easily shown that the solution of (\ref{23})
is given by
\begin{equation}
\label{24}
\frac{\phi(t)}{M_P} = \sqrt{ \frac{\gamma C}{\beta}}
\tanh \left[ -\sqrt{\frac{\beta\gamma}{C}} (t-t_i)
+ {\rm arctanh} \left(\frac{\phi_i}{M_P} \sqrt{ \frac{\beta}{C \gamma}}
\right) \right] + ...
\end{equation}
where
\begin{equation}
\label{gam}
\gamma \equiv -\alpha +\frac{\beta}{C}\left( \frac{\phi_i}{M_P} \right)^2.
\end{equation}
Using expression of Eq.(\ref{24}) in terms of the inverse powers of $C$,
one can obtain
\begin{equation}
\label{24a}
\frac{\phi(t)}{M_P} = \frac{\phi_i}{M_P} +
\alpha (t-t_i) + \frac{\alpha \beta}{C} (t-t_i)^2
\left[ \frac{\alpha (t-t_i)}{3} + \frac{\phi_i}{M_P} \right]
+ {\cal{O}} \left( \frac{1}{C^2} \right) + ...
\end{equation}
In the limit of constant $D$-space dimension, i.e. $C \to +\infty$,
we have 
\begin{equation}
\label{24b}
\frac{\phi(t)}{M_P} =
\frac{\phi_i}{M_P} - \frac{m D_0}{2 D_i} \sqrt{ \frac{D_i-1}{2 \pi D_i} }
(t-t_i).
\end{equation}
In this limit, we have $D=D_i=D_0$, and Eq.(\ref{24b}) can be rewritten
\begin{equation}
\label{24c}
\frac{\phi(t)}{M_P} =
\frac{\phi_i}{M_P} - \frac{m}{2} \sqrt{ \frac{D-1}{2 \pi D} }
(t-t_i).
\end{equation}
Also, the dynamical solution of the scale factor in the constant
$D$-space is given by \cite{1}
\begin{equation}
\label{24d}
a(t)=a_i \exp \left( \frac{4 \pi}{M_P^2(D-1)}[\phi_i^2-\phi^2(t)] \right).
\end{equation}
In the variable space dimension, to obtain the dynamical solution of
the scale factor, we rewrite Eq.(\ref{17})
\begin{equation}
\label{25}
\frac{\dot{a}}{a}=\frac{2 m \phi}{M_P}\sqrt{\frac{2\pi}{D(D-1)}}.
\end{equation}
We suggest that the solution of Eq.(\ref{25}) is analogous to
that of in the constant $D$-space. This means that we have
an exponential behavior for the scale factor
\begin{equation}
\label{26}
a(t)=a_i \exp \left( {\cal{F}} [\phi_i^2-\phi^2(t)] \right),
\end{equation}
where ${\cal{F}}$ is a function of $D, D_i$, and $D_0$.
To obtain ${\cal{F}}$, we differentiate Eq.(\ref{26}),
and use Eqs.(\ref{22}) and (\ref{25}). Therefore, one can get
\begin{equation}
\label{27}
{\cal{F}}=
\frac{8 \pi D}{D_0 M_P^2 \left[ \left( 2 - \frac{1}{D_i}
\right) D -1 \right]}.
\end{equation}
So, the solution of the scale factor in the chaotic inflation with
variable space dimension is given by
\begin{equation}
\label{28}
a(t)=a_i \exp \left( \frac{8 \pi D [\phi_i^2-\phi^2(t)]}
{D_0 M_P^2 \left[ \left( 2 -\frac{1}{D_i} \right) D -1 \right]}
\right).
\end{equation}
During the inflationary epoch with constant space dimension,
$D$ is equal to $D_i$, and Eq.(\ref{28}) approaches
\begin{equation}
\label{24f}
a(t)=a_i \exp \left( \frac{4 \pi D_i}{M_P^2D_0(D_i-1)}
[\phi_i^2-\phi^2(t)] \right).
\end{equation}
More carefully, in the limit of constant $D$-space, we have $D=D_i=D_0$,
and Eq. (\ref{24f}) gives Eq. (\ref{24d}).

Let us now calculate the dynamical behavior of the space dimension.
Substituting Eq.(\ref{24}) in Eq.(\ref{21}), one can obtain
\begin{eqnarray}
\label{29}
D(t) &=& \Bigg( 1+  \Bigg\{ 1 -
\frac{2}{D_i} \Bigg( 1-\frac{1}{2D_i} \Bigg)
- \frac{8 \pi}{CD_0} \Bigg[ \Bigg( \frac{\phi_i}{M_P} \Bigg)^2 \nonumber\\
&-&\frac{\gamma C}{\beta} \tanh^2 \Bigg[ -\sqrt{\frac{\beta \gamma}{C}}
(t-t_i)+{\rm arctanh} \Bigg( \frac{\phi_i}{M_P}
\sqrt{\frac{\beta}{C\gamma}} \Bigg) \Bigg]
\Bigg]  \Bigg\} ^{1/2} \Bigg) \nonumber\\ 
&\times& \Bigg\{ \frac{2}{D_i} \Bigg( 1-\frac{1}{2D_i} \Bigg)
+ \frac{8 \pi}{CD_0} \Bigg[ \Bigg( \frac{\phi_i}{M_P} \Bigg)^2\nonumber\\ 
&-&\frac{\gamma C}{\beta} \tanh^2 \Bigg[ -\sqrt{\frac{\beta \gamma}{C}}
(t-t_i)+{\rm arctanh} \Bigg( \frac{\phi_i}{M_P}
\sqrt{\frac{\beta}{C\gamma}} \Bigg) \Bigg] \Bigg] \Bigg\}^{-1}.
\end{eqnarray}
Expanding Eq.(\ref{29}) in inverse powers of $C$, one
can easily obtain
\begin{equation}
\label{30}
D= D_i + \frac{2 D_0 m (t-t_i)}{C} \Bigg( \frac{m (t-t_i)D_0}{4D_i}-
\frac{\phi_i}{M_P}\sqrt{\frac{2\pi D_i}{D_i-1}} \Bigg) + {\cal{O}} \Bigg(
\frac{1}{C^2} \Bigg) + ...
\end{equation}
This equation gives $D=D_i$ in the limit $C \to +\infty$, corresponding
to the constant space dimension, $D=3$.

\subsection{How long does the inflationary epoch last?}

Let us now obtain how long the inflationary epoch lasts.
Using Eqs. (\ref{19}) and (\ref{18h}), one can write

\begin{equation}
\label{31}
t-t_i=-\frac{\sqrt{C}}{m}
\int_{D_i}^D {\frac{dD}{\sqrt{\frac{D_0D(1-2D)}{(D-1)}+
\left[\frac{8\pi}{C}
\left(\frac{\phi_i}{M_P}\right)^2-\frac{D_0(1-2D_i)}{D_i^2}\right]
\frac{D^3}{(D-1)}}}},
\end{equation}
where the inflaton mass is $m \simeq 1.21 \times 10^{-6} M_P$ \cite{7}.
Inflation begins when $t=t_i$, and ends when $t=t_f$.
Integrating from $D_i$ to $D_f$, and using Table 1 and Table 2,
one can easily calculate how long the inflationary epoch lasts. This means 
\begin{eqnarray}
\label{32}
t_f-t_i &=& \cases {1.003 \times 10^{-36} \sec,          
               & if $\;D_P=4$, \cr
               1.159 \times 10^{-36} \sec,            
               & if $\;D_P=10$. \cr}
\end{eqnarray}
As given in \cite{1}, using Eq. (\ref{24c}) for $D=3$ one can obtain
$t_f-t_i$ in the case of constant space dimension
\begin{eqnarray}
\label{32a}
t_f-t_i&=&\frac{2 \sqrt{2\pi D} (\phi_i-\phi_f)}{m M_P
\sqrt{D-1}}\nonumber\\
& \simeq & 7.598 \times 10^{-37} \sec.
\end{eqnarray}
Comparing these values, one can conclude that the variability of
the space dimension causes the time duration of the inflationary epoch
to be longer than that of in the constant space dimension.
More carefully, when $D_P=4$ the value of $t_f-t_i$ is about 1.320 times
when the space dimension is constant
\begin{equation}
\frac{1.003 \times 10^{-36}}{7.598 \times 10^{-37}} \simeq 1.320.
\end{equation}
Also, when $D_P=10$ the value of $t_f-t_i$ is about
1.525 times when the space dimension is constant
\begin{equation}
\frac{1.159 \times 10^{-36}}{7.598 \times 10^{-37}} \simeq 1.525.
\end{equation}
So, the characteristic of the variability of the space dimension
causes the inflationary epoch in the variable space dimension
to be longer than the inflationary epoch in the constant space dimension.

As mentioned before, we do not consider $D_P=25$ or $+\infty$ in the model,
because for these values of $D_P$, the radius of th Universe at the
onset of inflation is smaller than the minimum length of the model,
i.e. $a_i < \delta$.
For this reason, in Eq.(\ref{32}) we have not calculted the value of
$t_f-t_i$ for $D_P=25$ or so. Contrary to the result of Ref.\cite{1},
we do not rule out the case of $D_P=10$ for which $a_i < l_P$.
For this value of $D_P$, we should consider the trans-Planckian problem
in inflationary cosmology with variable space dimension \cite{f}.

\section{CONCLUSIONS}

In this paper, we study the dynamical behavior of the inflaton field,
variable space dimension, and scale factor, in the framework of
a model Universe with variable space dimension proposed in \cite{4}.
Expanding the dynamical equations in terms of the inverse powers
of the universal constant of the model, we calculated the dynamical
solutions. Our study is based on the inflationary potential to be
equal to $m^2 \phi^2/2$.

We also show that chaotic inflation with variable space dimension
lasts longer than chaotic inflation with constant space dimension.
More carefully, the characteristic of the variability of the
space dimension causes the inflationary epoch of the model to last
longer than that of the standard cosmology in $3$-space.
In cosmology in $3$-space, chaotic inflation with the potential
$m^2 \phi^2/2$ lasts about $\sim 10^{-37}$ sec, while 
chaotic inflation with variable space dimension lasts about
$\sim 10^{-36}$ sec.

In this paper, we clear the meaning of the scale factor in the flat
Universe. Taking the comoving length of the horizon to be equal to one,
we consider the scale factor to be equal to the physical length of
the horizon.

Our treatments are based on the space dimension at the Planck length
to be equal to $4$ or $10$. As mentioned in \cite{1,3}, we rule out the
cases for which the space dimension at the Planck length to be equal to
$25$ or so, because in these cases the radius of the Universe at the
onset of inflation is smaller than the minimum size of the model.

Contrary to the result of Refs.\cite{1,3}, we do not rule out the case
for which the space dimension at the Planck length to be equal to $10$.
For this case, the radius of the Universe at the onset of inflation
is smaller than the Planck length, and one may consider
the inflationary cosmology with the variable space dimension in 
the trans-Planckian problem, which is a currently active
research [10-22]. We will address in detail this problem in \cite{f}.

\section*{Acknowledgments}

The author is grateful to an anonymous referee for useful comments.

\end{document}